\documentclass[structabstract]{aa}
\usepackage{txfonts}
\usepackage[utf8]{inputenc}
\usepackage[german,english]{babel}
\usepackage{natbib}
\usepackage{graphicx}
\usepackage{tabulary}
\usepackage{longtable}
\usepackage{lmodern}
\usepackage{rotating}
\usepackage{fp} 
\usepackage{url}
\usepackage{amstext}
\newcommand{\numoneall}{88330 }
\newcommand{\numone}{64172 }
\newcommand{\numtwo}{63958 }
\newcommand{\numthree}{60177 }
\newcommand{\numfour}{35261 }
\newcommand{\Minone}{1.066}
\newcommand{\Mintwo}{1.066}
\newcommand{\Minthree}{1.066}
\newcommand{\Minfour}{1.058}
\newcommand{\siginone}{0.094}
\newcommand{\sigintwo}{0.094}
\newcommand{\siginthree}{0.094}
\newcommand{\siginfour}{0.089}
\newcommand{\Moutone}{1.042}
\newcommand{\Mouttwo}{1.039}
\newcommand{\Moutthree}{1.018}
\newcommand{\Moutfour}{0.843}
\newcommand{\sigoutone}{0.160}
\newcommand{\sigouttwo}{0.168}
\newcommand{\sigoutthree}{0.216}
\newcommand{\sigoutfour}{0.394}
\newcommand{\Mdiffone}{0.032 }
\newcommand{\Mdifftwo}{0.033 }
\newcommand{\Mdiffthree}{0.036 }
\newcommand{\Mdifffour}{0.044 }
\newcommand{\sigdiffone}{0.058}
\newcommand{\sigdifftwo}{0.059}
\newcommand{\sigdiffthree}{0.061}
\newcommand{\sigdifffour}{0.078}
\newcommand{\Mshear}{0.088}

\newcommand{\Malphaout}{-0.056}
\newcommand{\sigalphaout}{0.098}
\newcommand{\lowerPone}{3.3}
\newcommand{\higherPone}{2.9}

\newcommand{\degree}{^\circ}
\newcommand{\dom}{\Delta\Omega}

\FPeval{\Pone}{round((\Minone-\Moutone)*100,1)} 
\FPeval{\secondper}{round(\numone/1000,1)} 
\FPeval{\Malpha}{round(\Mdiffone*100,1)}
\FPeval{\sigalpha}{round(\sigdiffone*100,1)}
\FPeval{\totshear}{round(\Mshear*100,1)}
\FPeval{\oneper}{round(96.2-\numone/1000,1)}

\begin{document}
\title{A fast and reliable method to measure stellar differential rotation from photometric data}

\author{Timo Reinhold \inst{1} \and Ansgar Reiners \inst{1} }

\offprints{T. Reinhold, \\ \email{reinhold@astro.physik.uni-goettingen.de} }

\institute{Institut für Astrophysik, Universität Göttingen, Friedrich-Hund-Platz 1, 37077 Göttingen, Germany}

\date{Received day month year / Accepted day month year}

\abstract
{
Co-rotating spots at different latitudes on the stellar surface generate periodic photometric variability and can be useful proxies to detect Differential Rotation (DR). DR is a major ingredient of the solar dynamo but observations of stellar DR are rather sparse. In view of the Kepler space telescope collecting more and more data we are interested in the detection of DR using photometric information of the star.
}
{
The main goal of this paper is to develop a fast method to determine stellar DR from photometric data.
}
{
We ran a large Monte-Carlo simulation of differentially rotating spotted stars with very different properties to investigate the detectability of DR. For different noise levels the resulting light curves are prewhitened using Lomb-Scargle periodograms to derive parameters for a global sine fit to detect periodicities.
}
{
We show under what conditions DR can successfully be detected from photometric data, and in which cases the light curve provides insufficient or even misleading information on the stellar rotation law. In our simulations, the most significant period $P1_{out}$ is on average \Pone\% lower than the actual spot rotation rate. This period could be detected in 96.2\% of all light curves. Detection of a second period close to $P1_{out}$ is the signature of DR in our model. For the noise-free case, in \secondper\% of all stars such a period was found. Calculating the measured latitudinal shear of two distinct spots $\alpha_{out}$, and comparing it to the known original spot rotation rates shows that the real value is on average \Malpha\% lower. Comparing the total equator-to-pole shear $\alpha$ to $\alpha_{out}$ we find that $\alpha$ is underestimated by \totshear\%, esp. the detection of DR for stars with $\alpha < 6\%$ is challenging. Finally, we apply our method to four differentially rotating Kepler stars and find close agreement with results from detailed modeling.
}
{
The method we developed is capable of measuring stellar rotation periods and detecting DR with relatively high accuracy and is suitable for large data sets. We will apply our analysis to more Kepler data in a forthcoming paper.
}

\maketitle 

\section{Introduction}\label{intro}
The solar dynamo generates the magnetic field of the Sun and causes its 11 year activity cycle. One of its major ingredients is believed to be Differential Rotation (hereafter DR) of the surface. The interaction of rotation and convection produces turbulence in the convection zone leading to non-uniform rotation \citep{Kitchatinov2005}. On the sun the angular velocity $\Omega$ decreases from the equator to the poles. Assuming an initial poloidal magnetic field with frozen field lines DR winds up the lines transforming an initial poloidal field into a toroidal field ($\Omega$-effect). The opposite effect which transforms the toroidal field back into a poloidal one is called $\alpha$-effect resulting from surface convection. In contrast to the $\Omega$-effect the $\alpha$-effect is able to produce a toroidal field from a poloidal one, and vice versa. Furthermore, the strength of DR varies with spectral type. \citet{Barnes2005} found that $\dom$ strongly increases with effective temperature supplying the power law $\dom\sim T_{\text{eff}}^{8.92}$. For temperatures above 6000 K this trend was confirmed by \citet{Reiners2006}. \citet{Cameron2007} combines results from Doppler Imaging (DI) and the Fourier transform method (see below) yielding the equation $\dom=0.053\;(T_{\text{eff}}/5130)^{8.6}$. This could be a hint towards different dynamo mechanisms, but the final role of DR is still not understood. \\
The relation between rotation period and DR has been studied by several authors. \citet{Hall1991} finds that the relative horizontal shear $\alpha$ increases towards longer rotation periods. \citet{Donahue1996} confirm this trend finding $\Delta P \sim \left\langle P \right\rangle^{1.3\pm0.1}$, independent of the stellar mass. Using the Fourier transform method \citet{Reiners2003} also find that $\alpha$ increases with rotation period for F-G stars. This result has been confirmed by \citet{Ammler2012} compiling previous results and new measurements for A-F stars. \citet{Barnes2005} find the relation $\dom\sim\Omega^{0.15}$ showing the weak dependence of DR on rotation rate. We believe that the potential of high precision photometry (e.g. provided by Kepler) has not been fully exhausted for measuring DR, esp. for cooler stars. \\
Solar DR has theoretically been studied for a long time. \citet{1999A&A...344..911K} compute DR models for late-type (G2 and K5) stars. They find that the relative shear $\alpha$ increases with rotation period. They also show that $\alpha$ increases towards cooler stars. \citet{2005A&A...433.1023K} compute models for an F8 star and find weak dependence of the total horizontal shear on rotation period, which is confirmed by later studies for F, G and K stars \citep{2005AN....326..265K} showing that the dependence of the horizontal shear on temperature is much stronger. The latter result holds for different main sequence star models \citep{2007AN....328.1050K}. They find that above a temperature of 5500 K the strong temperature dependence of the horizontal shear ($\sim T_{\text{eff}}^{8.92}$) from \citet{Barnes2005} fits the model data reasonably well whereas below 5500 K the data lies far off the fit. Recent studies \citep{2011AN....332..933K} have shown that the temperature dependence of $\Delta\Omega$ cannot be represented by one single power law over the whole temperature range from 3800-6700 K. The weak dependence of $\Delta\Omega$ on rotation period is confirmed for different solar mass models. \citet{2011ApJ...740...12H} model DR of rapidly rotating solar-type stars and find that DR approaches the Taylor-Proudman state, i.e. that $\Delta\Omega/\Omega$ decreases with angular velocity as long as the rotation rate is above the solar value.
\citet{Browning2011} provides an explanation for the increase of DR towards cooler stars. As the magnetic field increases towards cooler stars, its back-reaction on flows driven by Lorentz-force increases. Due to conservation of energy magnetic energy increases whereas the differential rotation kinetic energy decreases. \\
DR can be measured in different ways. \citet{Ammler2012} summarize the results using the Fourier transform method that analyzes the shape of Doppler broadened spectral lines \citep{Reiners2002a}.
DI tracks the migration of individual active regions over time to draw conclusions about the stellar rotation law. This method has successfully been used, e.g. by \citet{1997MNRAS.291....1D,2002MNRAS.330..699C}.
DR can also be detected by analyzing the star's light curve, e.g. \citet{Hall1991}. An analytical spot model \citep{Budding1977,Dorren1987} is fit to the light curve of $\epsilon$ Eridani in \citet{MOST2006} accounting for different spot periods. The same light curve has been analyzed by \citet{Froehlich2007} using MCMC method to estimate the parameters in a Bayesian way. This approach has also been applied to the CoRoT-2 light curve in \citet{Froehlich2009}. \citet{Kipping2012} presents an updated version of the spot model from \citet{Budding1977,Dorren1987} accounting for DR and spot evolution operating faster than previous models. \\
Following another approach, \citet{Lanza1993} create light curves of spotted stars and detect different periods by taking the Fourier transform.  \citet{Lucianne2012} fit an analytical spot model to synthetic light curves of spotted stars to see whether the model can break degeneracies in the light curve, esp. accounting for the ability of determining the correct rotation periods, both in the presence and absence of DR. In this paper we follow the same approach: We ran a large Monte-Carlo simulation producing 100.000 light curves trying to cover a significant portion of the parameter space. All light curves are then analyzed using the Lomb-Scargle periodogram incorporating a global sine fit. This is a fast and easy tool to derive periods from photometric data. Thus, the fundamental idea is to determine the accuracy of this method in detecting DR.
This is inevitable since with the advent of space telescopes like CoRoT and Kepler it became possible to study stellar flux variations of thousands of stars at the level of milli-magnitude precision. Different effects like co-rotating spots on the surface, pulsations, spot evolution, or instrumental effects can introduce periodic variations in the light curves, too. Therefore, the accuracy of this method needs to be tested for spot induced signals alone. \\
The paper is organized as follows: In sec. \ref{Simulations} the routine MODSTAR is introduced that synthesizes the light curves for the Monte-Carlo simulation. Sec. \ref{perdet} describes the period detection based on the Lomb-Scargle periodogram, with special focus on prewhitening, the selection process of the periods, and the sample properties. The results are found in sec. \ref{results}, starting with the most significant period $P1_{out}$ of the star in sec. \ref{P1}. In sec. \ref{dr} the DR results are presented. In sec. \ref{kepler} we compare our method to previously published differentially rotating Kepler stars. The results are summarized in sec. \ref{summary} briefly discussing the model parameters and analysis method. 
\section{Model description}\label{Simulations}
Stellar rotation in the presence of active regions leads to photometric variability in the stars' light curve. MODSTAR is our basic routine that creates a model star to simulate the photometric signal of a rotating spotted star. The star is modeled as a sphere with a fixed resolution of the surface pixels and inclination of the rotation axis. The intensity $I$ and projected area of each surface element depends on the value of $\mu$ which is the cosine of the angle between the pixel's surface normal and the line of sight. A quadratic limb darkening law is used:
\begin{equation}
 I(\mu)=I_0 (1-a(1-\mu)+b(1-\mu)^2),
\end{equation}
with $I_0$ being the intensity at the star's center. With respect to the Kepler mission we used the values $a=0.5287$ and $b=0.2175$ from \citet{Claret2000} relating to solar-like stars in the V-band. \\
Active regions can be placed on the surface. We use circular spots with desired longitudes, latitudes, radii, and a fixed intensity contrast. For the intensity contrast between the spots and the quiet photosphere we use a value of 0.67 which is approximately the solar penumbra to photosphere contrast. For simplicity, the spot pixels have all the same contrast value, i.e. we neglect umbra and penumbra structure. The stellar flux is integrated over the whole surface by summing up the pixel intensities weighted by projected area. Since the star can be rotated the flux is calculated at each rotation step which produces a light curve.
To describe the implemented rotation law, we quantify the amount of shear by
\begin{equation}\label{alpha}
  \alpha=\frac{P_{pole}-P_{eq}}{P_{pole}},
\end{equation}
with $\alpha=0$ supplying rigid body rotation. $P_{pole}$ and $P_{eq}$ are the rotation periods at the pole and the equator, respectively. $\alpha>0$ means that the equator rotates faster than the poles (solar-like DR) whereas $\alpha<0$ describes the opposite effect (anti solar-like DR). For our simulations, we only consider solar-like DR since one cannot discriminate between both effects exclusively from the light curve. The rotation period of a spot centered at a certain latitude $\theta$ is given by a common solar-like DR law:
\begin{equation}\label{spotper}
  P_{spot}(\theta)=\frac{P_{eq}}{1-\alpha\sin^2(\theta)}.
\end{equation}
According to eq.(\ref{spotper}) a spot would be torn apart after some rotation cycles. To avoid this the spots are fixed on the surface for all phases. In this way we achieve long-lived spots producing a stable beating pattern in the light curve. Evolution of spots is not included in our model. The spots are allowed to overlap with no further contrast reduction.

\subsection{Monte-Carlo simulation}
The Kepler mission provides light curves of all kinds of stellar activity, esp. rotation-induced variability. In some stars DR has been detected \citep{Frasca2011, Froehlich2012}, and many other light curves exhibit similar patterns. Inspired by this potpourri of active stars we asked the question to what accuracy DR can be measured solely from photometry if we allow for different kinds of stellar properties and spot configurations. \\
We ran a Monte-Carlo simulation producing 100.000 light curves of spotted stars to account for a large fraction of possible realizations. The most important stellar parameters are the inclination $i$, the number of spots on the surface, and the amount of DR $\alpha$. All parameters are uniformly distributed with $\sin(i)\in[0,1]$, the number of spots between 1 and 10, and $\alpha\in[0,1/3]$. The inclination covers the whole parameter space from pole-on ($i=0\degree$) to edge-on ($i=90\degree$) view.
\begin{table}
  \caption{Stellar simulation parameters}
  \label{stelpar}
  \begin{center}
    \begin{tabular}[c]{lcc}
      \hline \hline 
      Parameter & Value & Distribution\\
      \hline
      number of stars & 100.000 & -\\
      Inclination [$^\circ$] & 0 -- 90 & $\sin(i)$ uniform\\
      $\alpha$ value & 0 -- 1/3 & uniform\\
      number of spots & 1 -- 10 & uniform\\
      Period [$P_{eq}$] & 1 -- 1.5 & eq.(\ref{spotper})
    \end{tabular}
  \end{center}
\end{table}
\begin{table}
  \caption{Spot simulation parameters}
  \label{spotpar}
  \begin{center}
    \begin{tabular}{lccc}
      \hline \hline
      Parameter & Value [$^\circ$] & Distribution \\
      \hline
      longitude & -180 -- 180 & uniform \\
      latitude & -90 -- 90 & uniform \\
      radius  & 2 -- 21 & uniform
    \end{tabular}
  \end{center}
\end{table}
The spot positions are chosen at random and the spot radii are between 2$\degree$ and 21$\degree$ (s. Tables \ref{stelpar} \& \ref{spotpar}). The number of spots is limited to 10 because the spot radii can be rather large. These two limits prevent the star from being completely covered with spots which would result in a darker and therefore cooler star. The smaller the spot radii the more pixels are needed in our model to resolve individual surface elements. We fixed the minimum spot radius to 2$\degree$ to limit the computational effort. $\alpha$ covers the parameter space from rigid body rotation ($\alpha=0$) up to $\alpha=1/3$ including the solar value of $\alpha_{\sun}=0.20$. According to eq.(\ref{spotper}) the period is a function of latitude with $P=1$ cycle at the equator to $P=1.5$ cycles at the poles. A larger $\alpha$ value, e.g. $\alpha=0.5$ would result in rotation periods of $P=1$ cycle at the equator to $P=2$ cycles at the poles. This would lead to problems in discriminating between harmonics and true rotation rates. In observations values of $\alpha>0.50$ have been found \citep{Ammler2012}. DI usually yields lower values of the order $\alpha\lesssim 0.01$, see Table 1 in \citet{Barnes2005}, and references therein. Our method is sufficient to produce light curves of differentially rotating stars for a wide range of $\alpha$ but becomes problematic for either very high or very low $\alpha$ values. This issue is discussed in sec. \ref{summary}. \\
The spot to photosphere contrast was fixed because there is a degeneracy between spot size and contrast. Choosing the contrast as free parameter in the simulation would just change the depth of the spot signature in the light curve but does not affect the period. The limb darkening coefficients are fixed for all light curves. Each light curve consists of 300 data points covering 10 rotation periods $P_{eq}$ to see how the light curves evolve in time. \\
In the following we consider different noise levels in the light curves: the noise-free case, 100 ppm, 1000 ppm, and 10.000 ppm Poisson noise which is added to the light curves. A minimum noise level of 100 ppm is chosen because it is lower than the depth produced by the smallest spot (2$^\circ$) which we find to be approx. 250 ppm. The same argument applies to the largest noise level which is lower than the depth of the largest spot (21$^\circ$) being approx. 48.000 ppm. 
An example of a 1000 ppm model light curve is shown in Fig. \ref{lc130_fit}. This light curve looks similar to what we see in Kepler data so we are optimistic that the parameter selection is sufficient for our main purposes -- the production of a variety of light curves with periodic variability and the detection of DR. We want to keep the model simple and try to see whether this can reproduce real light curves. A larger parameter space can be tackled if we find that our model cannot reproduce the data. A similar argument applies to spot evolution, which so far we separate from our approach.

\section{Period determination}\label{perdet}
We detect DR by measuring different periods in a light curve. Since we are facing a large sample a fast and reliable frequency analysis tool is needed. We chose the Lomb-Scargle periodogram which is widely used in time series analysis. Calculating the periodogram of a single light curve takes only one second. Although being a purely mathematical tool the program is sufficient to find different periods in the data. Fitting an analytical spot model also supplies rotation periods and several other stellar parameters but severely slows down the analysis process.

\subsection{Lomb-Scargle periodogram}\label{LS}
The Generalised Lomb-Scargle periodogram \citep{Zechmeister2009}\footnote{For different periodogram codes see \url{http://www.astro.physik.uni-goettingen.de/~zechmeister/gls.php}} is a powerful spectral analysis tool for unevenly sampled data. It fits the data using a series of sines and cosines. The frequency grid used for the fit is sampled equidistantly. Its range has an upper limit due to the Nyquist frequency. The lower limit is given by the inverse product of the time span and a desired oversampling factor to achieve a proper frequency resolution. We use a factor of 10, i.e. a minimum frequency of 0.01/cycle. Depending on the goodness of the fit one obtains peaks with different powers -- the better the fit, the higher the peak in the periodogram. The periodogram is normalized to unity. The period P (or frequency f=1/P) associated to the highest peak is the most dominant one in the data. In some cases an alias of P/2, P/3, etc. may produce a peak with high power, too. One reason for a higher alias than the rotational period one can be the presence of two active longitudes separated approx. $180\degree$ from each other. Another one can be the improper shape of a sine wave to fit the spot signature. A single spot does not produce a sinusoidal shape except for pole-on view. In frequency domain aliases are equidistant (2f, 3f, etc.) and can easily be detected by eye because the peak height usually decreases towards higher harmonics. We get rid of most alias periods as described in sec. \ref{selection}.

\subsection{Prewhitening}\label{pw}
\begin{figure}
  \resizebox{\hsize}{!}{ \includegraphics{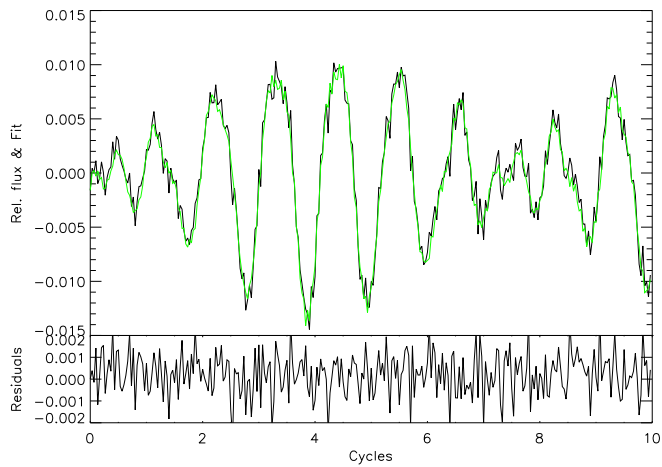} }
  \caption{\textit{Upper panel}: Model light curve with 1000 ppm Poisson noise added and the best global sine fit over-plotted (green). The spot periods of the model and the ones returned by our method are given in Table \ref{table_130}. \textit{Lower panel}: Residuals of the best fit subtracted from the data; no periodicity visible anymore.}
  \label{lc130_fit}
\end{figure}

In most cases our model stars are covered by several spots adding up their signals to one light curve. Thus, we are facing the challenge to fit a mixture of period signatures. These periods are extracted from the light curve in a successive way called prewhitening. First, we adopt the period associated to the highest peak in the periodogram and fit a sine wave to the light curve. The initial sine function is subtracted from the data and another periodogram is taken from the residuals. Again, we fit a sine function and subtract it from the data. This prewhitening process can be repeated as often as desired, i.e. until there is no periodicity present anymore. On the one hand, a high number of prewhitening steps is crucial for the detection of several periods, but on the other hand, prewhitening is computationally intensive and one has to be careful to select the correct periods afterwards (sec. \ref{selection}) which becomes more difficult with a larger set of periods. Since the stars are covered by 10 spots or fewer we repeat this procedure 10 times for each light curve. Finally, all 10 periods detected during prewhitening are used as input parameters for a global sine fit, which is the sum of 10 sine functions with different periods, amplitudes, phases, and one total offset. The result of this last step is an optimal set of parameters found through $\chi^2$ minimization. A model light curve with 1000 ppm noise and the fit obtained through this procedure are displayed in Fig. \ref{lc130_fit}.
In the first column of Table \ref{table_130} the actually contained spot periods are shown; this particular example has only 3 spots. The second column contains all detected periods. The latter ones are arranged as follows: The first period is the one with the highest power found in the prewhitening process, the second one belongs to the second highest power, and so on. We see that in this case the third and fifth period are harmonics of the first and second one, resp., the fourth one and the last five periods just fit the remaining noise. The fit in Fig. \ref{lc130_fit} shows good agreement with the light curve and the residuals carry no more periodicity in the domain of one or more cycles. 

\subsection{Period selection}\label{selection}
The fitting procedure described above returns 10 periods for each light curve. In this section we assign a physical meaning to the most significant period (comp. Table \ref{table_130}), and show how to detect further periods close to it as evidence for DR. Even though we know that in our model the spot periods range from 1-1.5 cycles (eq.\ref{spotper}), we will not constrain our algorithm to this range since in reality we don't know the correct period either. \\
The first sine period is the most significant one in the data. In several cases this period is equal to the first harmonic (P/2) of the true spot rotation rate due to certain spot configurations, e.g. two spots located on opposite sides of the star. In these cases the second sine period is likely the correct period P. To minimize the number of alias periods we compare the double of the first period to the second sine period. If these two differ by less than 5\% then the second sine period is chosen. The finally selected period is our primary period $P1_{out}$. In sec. \ref{P1} we show that this period yields the best estimate of the stellar rotation rate. In most cases the light curves result from several spots on the surface so we need to compare $P1_{out}$ to the spot periods according to eq.(\ref{spotper}). This set is called the \textit{input periods} since they belong to spots inherent in the light curves. Thereof, we take the one closest to $P1_{out}$ and call it $P1_{in}$. \\
Based on the period $P1_{out}$ we looked for a second period which we call $P2$ for the moment. In order to attribute this period to a second spot one has to balance three things: 1) Find a second period that is no harmonic of $P1_{out}$, 2) try to exclude as few spot periods as possible, and 3) try to dismiss all period artifacts that come from fitting a sine wave to the light curve rather than a spot model. Therefore, $P2$ should hold the relations
\begin{equation}\label{limits}
  0.01 \leq \frac{|P1_{out}-P2|}{P1_{out}} \leq \alpha_{max},
\end{equation}
with $\alpha_{max}=1/3$. The value $\alpha_{max}=1/3$ corresponds to the maximum $\alpha$ in our model. As mentioned above a higher value near $\alpha_{max}\approx0.5$ would yield ambiguous results for the DR. Image two cases: 1) $P1_{out}=1, P2=0.5$, and 2) the case $P1_{out}=1, P2=1.5$. In both cases a second period would be selected but due to completely different origins. In case 1) the first harmonic of $P=1$ would be mis-interpreted as DR, whereas case 2) results from real spot configurations. We chose the value $\alpha_{max}=1/3$ because it excludes harmonics and covers a wide range for a second spot period. For example, in the extreme case of $P1_{out}=1$, the harmonic at $P2=1/2$ is excluded but we are not able to find a spot period greater than $P=1.33$ although there might be spots with longer periods. The lower limit in relation (\ref{limits}) accounts for the fixed frequency resolution in the Lomb-Scargle periodograms (s. sec. \ref{LS}). If two spot periods differ by less than 1\% they cannot be resolved. Again, we might miss some spot periods lying closer than 1\% with our method. \\
If one or more periods were found fitting both criteria (compare Table \ref{table_130}) we took that $P2$ associated to the period with the lowest row index in the table of remaining\footnote{If $P1_{out}$ was a harmonic then the first two sine periods are excluded.} sine periods. This period is called $P2_{out}$. In \numoneall cases of our models we find a $P2_{out}$ that fulfills these criteria. Again, $P2_{out}$ is compared to all input periods and the period closest to $P2_{out}$ is called $P2_{in}$. If the closest input period picked is again $P1_{in}$ then $P2_{out}$ is discarded. Finally, we are left with \numone stars having two periods which belong to two different spots. \\
For our example light curve in Fig. \ref{lc130_fit}, the periods $P1_{out}=1.016, P2_{out}=1.184, P1_{in}= 1.018$, and $P2_{in}= 1.186$ have been selected from Table \ref{table_130}, resulting in $\alpha_{in}=0.14$ and $\alpha_{out}=0.14$. Although our method is able to recover the correct $\alpha_{in}$ value, the total equator-to-pole shear equals $\alpha=0.30$ in this case (comp. eq.(\ref{spotper})), and thus is underestimated by more than 50\%. For this specific spot configuration it is impossible to obtain the correct $\alpha$ value since the highest spot latitude equals $\theta=46.3\degree$ generating the longest period. This is a general problem of DR measurements from photometric observations due to the initial spot configuration on the surface. The measured shear will always yield a lower value than the total one.
\begin{table}
  \caption{\textit{Left}: Periods of the three spots from the light curve in Fig. \ref{lc130_fit} (left column) and output periods returned by the prewhitening analysis for the fit in Fig. \ref{lc130_fit} (right column). \textit{Upper right}: The two periods $P_{in}$ and $P_{out}$ that have been selected from the left table. \textit{Lower right}: The resulting values $\alpha_{in}$ and $\alpha_{out}$ computed from the upper table. The total equator-to-pole shear equals $\alpha=0.30$ in this case which is underestimated by more than 50\%.}
  \label{table_130}
  \begin{center}
    \hspace{-3.5cm}
    \begin{tabular}{cc}
$P_{spot}$ & $P_{out}$ \\
\hline
1.049 & 1.016 \\
1.018 & 1.184 \\
1.186 & 0.510 \\
-     & 0.906 \\
-     & 0.591 \\
-     & 0.078 \\
-     & 0.081 \\
-     & 0.340 \\
-     & 0.079 \\
-     & 0.134 
\end{tabular}

    \put(8,40){$\Longrightarrow$}
    \put(30,11){\begin{tabular}{cc}
  $P_{in}$ & $P_{out}$ \\
  \hline
  1.018 & 1.016 \\
  1.186 & 1.184 \\
  \\
& \hspace{-1.3cm}
\begin{turn}{90}$\Longleftarrow$\end{turn} \\
  \\
  $\alpha_{in}$ & $\alpha_{out}$ \\
  \hline
  0.14 & 0.14
\end{tabular}
}
    \tablefoot{Period selection process using the example of the light curve from Fig. \ref{lc130_fit}.}
  \end{center}
\end{table}

\subsection{Sample Properties}\label{compare}
\begin{figure*}
 \centering
 \includegraphics[width=17cm]{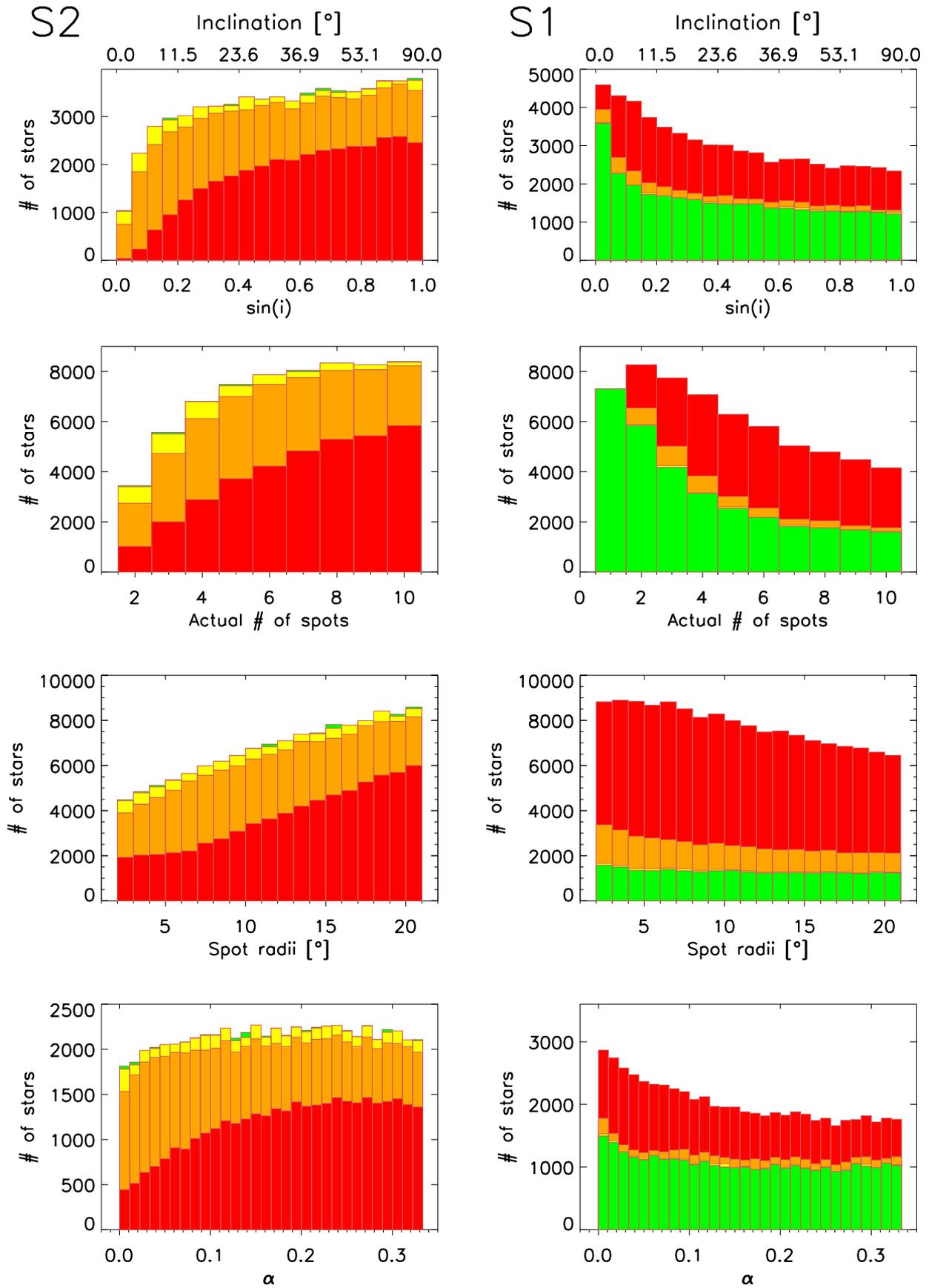}
 \caption{Comparison of basic stellar parameters for both samples S2 and S1 for different noise cases: noise-free (green), 100 ppm (yellow), 1000 ppm (orange), and 10.000 ppm (red) Poisson noise. \textit{Left panel}: Stars with two spot periods (S2). \textit{Right panel}: Stars with only one spot period (S1).}
 \label{comp}
\end{figure*}

This section is thought as a consistency check of our model and the selection algorithm. The stellar parameters (s. Table \ref{stelpar}) of two mutually exclusive samples, S2 and S1, are compared. The S2 sample consists of all light curves with two detected periods (\secondper\%) coming from two distinct spots, whereas S1 contains all cases where only one spot period could be associated (\oneper\%). Due to a combination of a low number of spots, certain spot latitudes, and a low inclination in 3.8\% of all cases no spot was visible. \\
The goal is to point out those stellar models where DR can likely be found compared to the cases where the detection of DR is challenging or even impossible with our method. For example, one would expect that it is easier to detect DR in the case of a highly spotted star rather than in the case of a star covered by only two close-in spots because it will probably be hard to resolve individual periods in the latter case. \\
In Fig. \ref{comp} we compare the inclination, number of spots, spot radii, and differential rotation $\alpha$ of both samples. The S2 sample is shown in the left and the S1 sample in the right column, resp.. The colors correspond to the different noise levels: noise-free (green), 100 ppm (yellow), 1000 ppm (orange), and 10.000 ppm (red) Poisson noise.
Due to the small difference of detections for the noise-free (green) and the 100 ppm noise case (yellow) there is basically no difference visible in the histograms. All trends have similar shape and become more distinct towards higher noise levels. \\
In the first row the inclination for both samples is plotted. Remember that $\sin(i)$ of the whole sample has a flat distribution. Starting from edge-on view ($i=90\degree$) we see a continuous decrease (increase) of detections in the S2 (S1) sample. Around inclinations lower than $i=10\degree$ the number of detections in S2 decreases significantly. The opposite effect applies to S1. \\
In the second row we show distributions of the actual number of spots of the models. We find that only in a very few cases the models in S2 can be attributed to only two spots on the surface. In the majority of all cases the light curve is composed by the signature from more than 5 spots! We find that the actual number of spots decreases in the S1 sample. \\
The third row shows the distribution of the spot radii. In the left panel one clearly sees that the radii of both spots found increases to higher values because it is more likely to detect more than 1 period if the star has large spots. In the right panel we plot the radii of all visible spots except for the one associated to the one period found. These "residual`` or not resolved spots show a shallow decrease in radii. \\
Finally, the last row shows histograms of the $\alpha$ value. Both distributions are basically flat except for the 10.000 ppm noise case and for small values of $\alpha$. The distribution of S2 decreases while the one of S1 increases towards lower $\alpha$ values. \\
All above histograms show consistent results for each sample supporting the underlying model. With a focus on the S2 sample the selection process seems convincing to pick mostly those models where the detectability of DR is expected. The derived periods and accuracy of our tool are discussed in the following section.

\section{Results}\label{results}
In this section we compare the outcome of our analysis to the periods from our model. First, we present the basic results for the most dominant period $P1_{out}$. In sec. \ref{dr} the detection of DR is discussed considering different noise levels. Finally, we compare our method to three Kepler stars where DR has been confirmed (sec. \ref{kepler}).

\subsection{Rotation periods $P1_{out}$}\label{P1}
\begin{figure*}
  \centering
  \includegraphics[width=17cm]{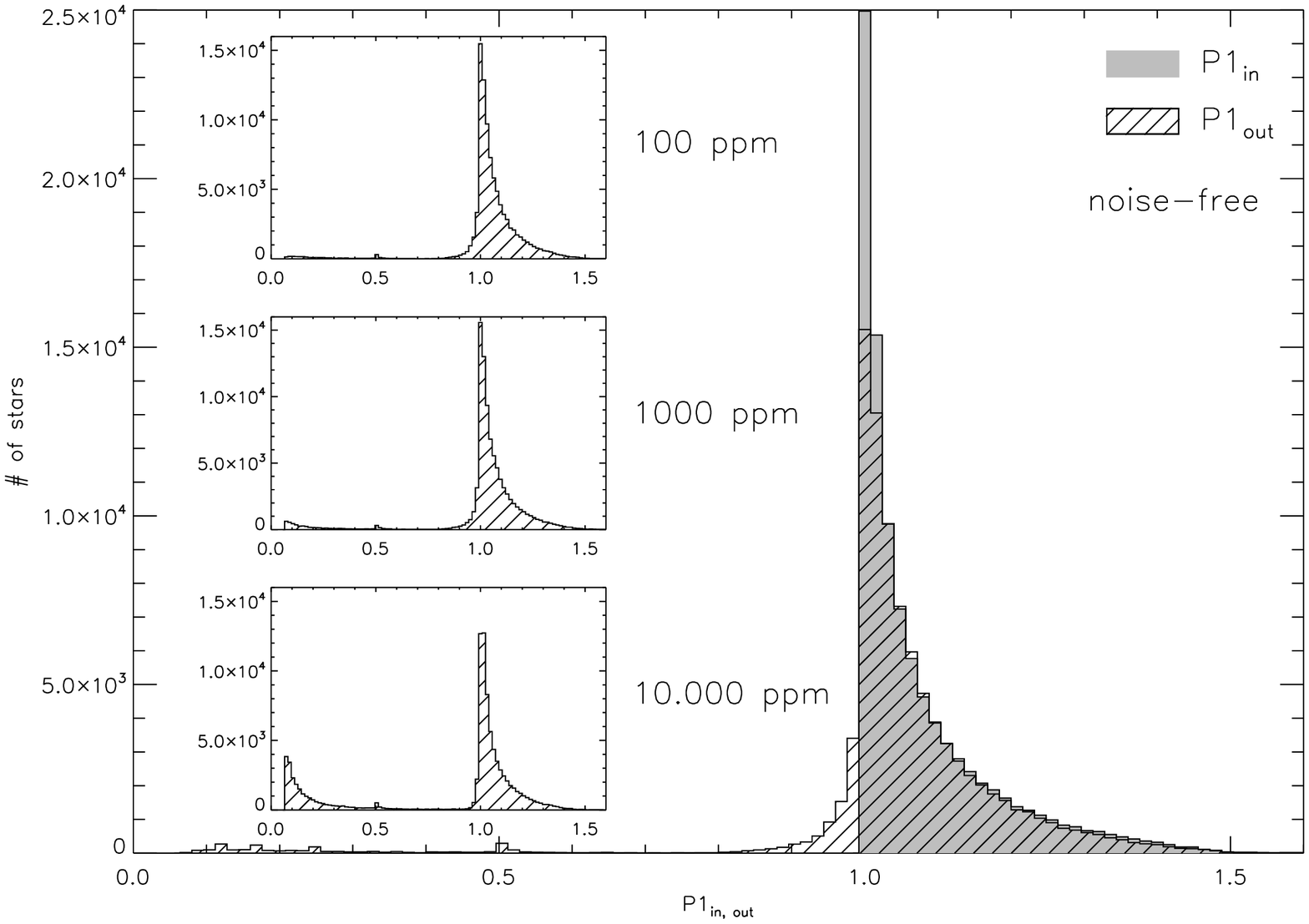}
  \caption{\textit{Main plot:} Comparison of exactly known input period $P1_{in}$ and output period $P1_{out}$ for the noise-free case. We see that the most significant period $P1_{out}$ can be recovered fairly well. We find that $P1_{out}$ is on average \Pone\% lower than the actually contained period $P1_{in}$. \textit{Small plots:} The distribution of $P1_{out}$ is shown with increasing noise from top to bottom. Towards higher noise levels the fraction of periods lower than 0.5 cycles increases because the algorithm interprets the noise as short periods (comp. Table \ref{rot_table}).
	  }
  \label{P1_eq}
\end{figure*}
In 96.2\% of all light curves there is one detected periodicity $P1_{out}$ which is the most significant one in the data. In Fig. \ref{P1_eq} we compare the periods $P1_{out}$ to the input periods $P1_{in}$ for different noise levels to see how good the above selection process works. The distribution of input periods $P1_{in}$ is shown in gray, and the shaded distribution shows the output periods $P1_{out}$ . Since we only consider solar-like DR $P1_{in}$ cannot be lower than 1 cycle. The shaded gray area shows that both histograms overlap quite well. For the noise-free case the number of wrong detections is negligible with \lowerPone\% of all periods being lower 0.9 cycles and \higherPone\% being greater than 1.6 cycles. This is no longer true for higher noise levels. The region from 0 - 0.5 cycles becomes populated as shown in the 3 small plots in Fig. \ref{P1_eq}. We compare the weighted means $\left\langle P1_{in}\right\rangle$ and $\left\langle P1_{out}\right\rangle$ in the above range in Table \ref{rot_table}. In the noise-free case we find that $P1_{out}$ is on average \Pone\% lower than the actually contained period $P1_{in}$. For higher noise levels $\left\langle P1_{out}\right\rangle$ decreases because the algorithm interprets the noise as short periods. In the noise-free case the wrong detections are due to a low stellar inclination close to pole-on view or in some cases due to higher harmonics. The cases where $P1_{out}$ is lower than 1 cycle are due to the improper shape of a sine function to fit spot signatures in a light curve. For example, a detected period of $P1_{out}=0.98$ cycles will be considered as valid rotation period although it is not possible for the spots to rotate this fast in our model. This fact will not be noticed in real data because we do not have information on the real rotation of a star. Around a period of 0.5 cycles only a small fraction of harmonics remained (less than 0.5\%) after identification and correction (s. sec. \ref{selection}).
\begin{table}
  \caption{Weighted means $\left\langle P1_{in}\right\rangle$ and $\left\langle P1_{out}\right\rangle$, and their associated errors $\sigma(P1_{in})$ and $\sigma(P1_{out})$ for both input and output periods, resp. for each noise level.}
  \label{rot_table}
  \begin{center}
    \begin{tabular}{ccccc}
  \hline\hline
  Noise [ppm] & $\left\langle P1_{in}\right\rangle$ & $\sigma(P1_{in})$ & $\left\langle P1_{out}\right\rangle$ & $\sigma(P1_{out})$ \\
  \hline
	0 & \Minone & \siginone & \Moutone & \sigoutone \\
      100 & \Mintwo & \sigintwo & \Mouttwo & \sigouttwo \\
    1000 & \Minthree & \siginthree & \Moutthree & \sigoutthree \\
    10.000 & \Minfour & \siginfour & \Moutfour & \sigoutfour
\end{tabular}
  \end{center}
\end{table}

\subsection{Differential Rotation}\label{dr}
\begin{figure*}
  \centering
  \includegraphics[width=17cm]{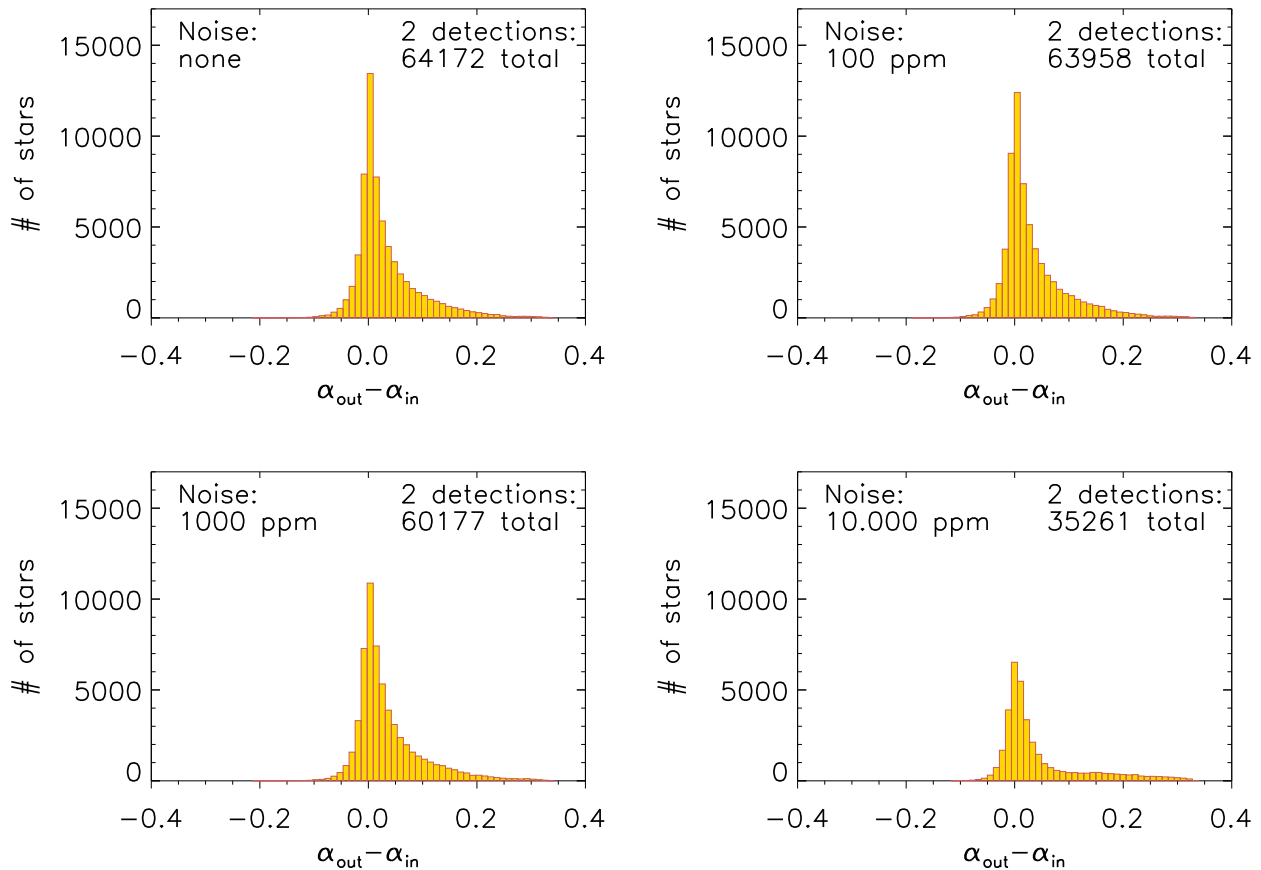}
  \caption{Histograms of $\alpha_{out}-\alpha_{in}$ for all noise levels. From the upper left to the lower right panel the noise increases. For the noise-free case (upper left panel) in \numone stars (at least) two periods are detected. The distribution is centered at $\left\langle\alpha_{out}-\alpha_{in}\right\rangle=\Mdiffone$ with a width of $\sigma(\alpha_{out}-\alpha_{in})=\sigdiffone$. Going to higher noise levels the total number of findings decreases and the error increases (s. Table \ref{alpha_table}).
	  }
  \label{alphadiff}
\end{figure*}

\begin{figure}
  \resizebox{\hsize}{!}{ \includegraphics{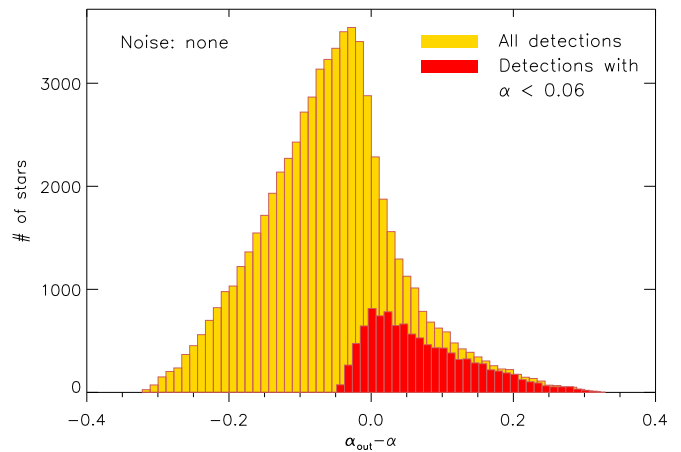} }
  \caption{The histogram shows the distribution of $\alpha_{out}-\alpha$ for the noise-free case. It is centered at $\left\langle\alpha_{out}-\alpha\right\rangle=\Malphaout$ with a width of $\sigma(\alpha-\alpha_{out})=\sigalphaout$. Comparing this plot to Fig. \ref{alphadiff} shows that the total amount of DR is underestimated by \totshear\% because $\alpha_{in}\leq\alpha$ by definition. The detections where $\alpha < 0.06$ are over-plotted in red. It is evident that in these cases the detection of DR is difficult. In most cases, $\alpha_{out} > \alpha$ which demonstrates the limits of our method
	  }
  \label{diff_alpha_out}
\end{figure}

In this section we show in which situations DR can successfully be detected and which cases lead to wrong interpretations. In our model the detection of a second period $P2_{out}$ adjacent to the primary one and associated to a second spot period $P2_{in}$ is considered as evidence for DR. To see where this selection process is acceptable we consider the two pairs ($P1_{out}, P2_{out}$) and their associated periods ($P1_{in}, P2_{in}$). We estimate the amount of DR by sorting these pairs and computing their $\alpha_{in, out}$ value, resp. (s. Table \ref{table_130}):
\begin{equation}
  \alpha_{in, out}=\left.\frac{P1-P2}{P1}\right|_{in, out}, P1>P2.
\end{equation}
The $\alpha_{in}$ value is always lower than the inherent $\alpha$ value of each light curve since we can only measure the rotational shear at two defined latitudes. If $\alpha_{in}$ is calculated from the spots with the largest separation in latitude on a certain hemisphere then $\alpha_{in}$ is the maximum shear that can be detected by our method. The distribution of $\alpha_{out}-\alpha_{in}$ gives a statistical measure of the robustness of our period selection process. In Fig. \ref{alphadiff} we show the distribution of the differences between $\alpha_{out}$ and $\alpha_{in}$ in our set of models for different noise levels. All distributions exhibit an asymmetric shape towards too large $\alpha_{out}$ values. An explanation for $\alpha_{out}$ being too large is given at the end of this section. \\
\begin{table}
  \caption{Weighted mean $\left\langle\alpha_{out}-\alpha_{in}\right\rangle$ and error $\sigma(\alpha_{out}-\alpha_{in})$ of $\alpha_{out} - \alpha_{in}$ for all stars with two detected periods for different noise levels.}
  \label{alpha_table}
  \begin{center}
    \begin{tabular}{cccc}
  \hline\hline
  Noise [ppm] & \# of stars & $\left\langle\alpha_{out}-\alpha_{in}\right\rangle$ & $\sigma(\alpha_{out}-\alpha_{in})$ \\
  \hline
	0 & \numone & \Mdiffone & \sigdiffone \\
      100 & \numtwo & \Mdifftwo & \sigdifftwo \\
    1000 & \numthree & \Mdiffthree & \sigdiffthree \\
  10.000 & \numfour & \Mdifffour & \sigdifffour
\end{tabular}
  \end{center}
\end{table}
The total number of light curves with two (or more) detected periods, the weighted mean, and the error of the distributions are given in Table  \ref{alpha_table}. In general, the total number of stars with two detected periods decreases with increasing noise. All differences between the noise-free and the 100 ppm case are marginal. The 1000 ppm case has slightly less detections, and the most significant decrease happens for the 10.000 ppm case (compare Fig. \ref{comp}). For all cases the weighted mean and the error increase with noise. Considering the noise-free distribution (upper left) we find a weighted mean $\left\langle\alpha_{out}-\alpha_{in}\right\rangle=\Mdiffone$ and a width of $\sigma(\alpha_{out}-\alpha_{in})=\sigdiffone$. The error is given in absolute units of $\alpha_{out}-\alpha_{in}$ regardless of the true shear $\alpha$. We also considered the relative errors $|\alpha_{out}-\alpha_{in}|/\alpha$ for $\alpha>0.05$. Each distribution is proportional to const./$\alpha$, i.e $\sigma(\alpha_{out}-\alpha_{in})$ does not scale with $\alpha$. For each noise level the number of stars decreases with increasing relative error. We only considered $\alpha>0.05$ to avoid large errors.
The statistical error resulting from the sample of the light curves used is negligible. The above calculations have been done for two additional sets of light curves (each 100.000 in total). We find that $\left\langle\alpha_{out}-\alpha_{in}\right\rangle$ and $\sigma(\alpha_{out}-\alpha_{in})$ are almost equal for each set, and that the largest statistical uncertainty in the number of stars with two detected periods is about 0.3\%. \\
We have shown that DR can be measured with high accuracy for a large noise range. In Fig. \ref{diff_alpha_out} we compare $\alpha_{out}$ to the total equator-to-pole shear $\alpha$ of the star.
In the noise-free case the distribution has a weighted mean $\left\langle\alpha_{out}-\alpha\right\rangle=\Malphaout$ and an error $\sigma(\alpha_{out}-\alpha)=\sigalphaout$. Considering the mean values yields
\begin{eqnarray*}
  \Mdiffone &=& \left\langle\alpha_{out}-\alpha_{in}\right\rangle \geq \left\langle\alpha_{out}-\alpha\right\rangle=\Malphaout \\
  &\Rightarrow& \left\langle\alpha_{out}-\alpha\right\rangle \leq 0.088
\end{eqnarray*}
since $\alpha_{in}\leq\alpha$ by definition. This means that the total amount of DR is underestimated by \totshear\% by our method. Models with small $\alpha$ values exhibit spot periods very close to each other. Thus, we tested whether these are prone to mis-identification of DR. In Fig. \ref{diff_alpha_out} the models with an equator-to-pole shear $\alpha<0.06$ are over-plotted in red. We find that $\alpha_{out}$ yields wrong values larger than $\alpha$ itself. This results from the difficulty to resolve two distinct peaks in the Lomb-Scargle periodogram. Since the peak width is proportional to the inverse time span of the light curve we are not able to resolve two periods within 10 cycles. One broadened peak appears resulting in a \textit{mean} period. In these cases the initial sine wave does not have a proper shape to fit the mixture of spot signatures so there remain artifacts which are corrected by fitting more sine waves. If one of the residual periods fulfills the selection criteria the algorithm selects it as $P2_{out}$ yielding too high $\alpha_{out}$ values. This behavior partly applies to the distributions in Fig. \ref{alphadiff}, esp. for the highest noise case. A second spot period cannot be resolved properly and is lost in the noise yielding an asymmetric distribution.

\subsection{Comparison to Kepler data}\label{kepler}
We have seen that our method performs well for simulated data. To test its reliability we applied it to three previously studied Kepler stars \citep{Frasca2011,Froehlich2012} where DR is the favorite explanation for the light curve variability. In both papers synthetic light curves from an analytical spot model \citep{Dorren1987} are fit to the data, and the parameters are estimated in a Bayesian way using MCMC methods. The stars KIC 8429280 \& KIC 7985370 are fit with 7 spots, and the star KIC 7765135 is fit with 9 spots. In their model each spot has a certain rotation period. Further model parameters are the equator-to-pole differential rotation d$\Omega$, and \citet{Froehlich2012} also use the equatorial period $P_{eq}$. We used our method to determine $P1_{out}$ and $P2_{out}$ to compare them to the reported results. We call the spot periods closest to our findings $P1_{closest}$ and $P2_{closest}$, resp.. From these periods we calculate $\alpha_{out}$ and $\alpha_{closest}$.The last row in Table \ref{kic_table} contains the calculated equator-to-pole shear d$\Omega/\Omega_{eq}$ of the models which equals $\alpha$ in our simulations. Since KIC 8429280 has no parameter for the equatorial period we used the one from the spot closest to the equator ($\theta=-0.5\degree$). For all three stars we determine consistent results for the horizontal shear (s. Table \ref{kic_table}). \\
Some spots fitted in the model have very close or even identical periods. According to relation (\ref{limits}) we are not able to detect periods closer than 1\%. Even without this restriction one would need a very high frequency resolution in the Lomb-Scargle periodograms to see these shallow differences. Thus, we think that the most significant periods in the data have been successfully detected by our method because the returned $\alpha$ values are very close. Recently, \citet{Roettenbacher2013} determine a surface rotation period of 3.47 days for the Kepler target KIC 5110407, and evidence for DR and spot evolution using light curve inversion. We find $P1_{out}=3.61$ and $P2_{out}=3.42$ days yielding $\alpha_{out}=0.052$. Their DR coefficient $k$ (equal to $\alpha$ in our case) strongly depends on the assumed inclination angle and varies between $k=0.024-0.118$. Using $i=45\degree$ they find $k=0.053$ which is very close to our result. 
Hence, our method successfully returned a consistent DR coefficient using a faster and much easier method than light curve inversion.

\begin{table}
  \caption{Comparison of periods returned by our method to spot periods from previously studied Kepler stars. The periods closest to $P1_{out}$ ($P2_{out}$) are called $P1_{closest}$ ($P2_{closest}$), resp..}
  \label{kic_table}
  \begin{center}
    \begin{tabular}{ccccc}
\hline\hline
KIC & 8429280 & 7985370 & 7765135 & 5110407 \\
\hline
$P1_{out}$ & 1.16194 & 2.84359 & 2.55393 & 3.61073 \\
$P2_{out}$ & 1.21012 & 3.09043 & 2.39581 & 3.42325 \\
$\alpha_{out}$ & 0.040 & 0.080 & 0.062 & 0.052 \\
\hline
$P1_{closest}$ & 1.16298\tablefootmark{a} & 2.8428\tablefootmark{b} & 2.5645\tablefootmark{c} & - \\
$P2_{closest}$ & 1.20430\tablefootmark{a} & 3.0898\tablefootmark{b} & 2.4018\tablefootmark{c} & 3.4693\tablefootmark{d} \\
$\alpha_{closest}$ & 0.034 & 0.080 & 0.063 & - \\
\hline
d$\Omega/\Omega_{eq}$ & 0.049 & 0.080 & 0.067 & 0.053\tablefootmark{d} \\
\hline
\end{tabular}

    \tablefoot{
    \tablefoottext{a}{Periods from 7-spots fit \citep{Frasca2011}.}
    \tablefoottext{b}{Periods from 7-spots fit \citep{Froehlich2012}.}
    \tablefoottext{c}{Periods from 9-spots fit \citep{Froehlich2012}.}
    \tablefoottext{d}{Period and DR coefficient from \citet{Roettenbacher2013}.}
    }
  \end{center}
\end{table}

\section{Summary}\label{summary}
We have run a large Monte-Carlo simulation of differentially rotating, spotted stars covering a significant fraction of the parameter space. The resulting light curves have been analyzed with the Lomb-Scargle periodogram in a prewhitening approach. The returned periods have been used for a global sine fit to the data. The major goal was to see under which stellar conditions and upon what accuracy DR can be detected. \\
The most significant periodicity $P1_{out}$ could be detected in 96.2\% of all light curves. A second period close to $P1_{out}$ has been attributed to a second spot found in \secondper\% of all cases. The latitudinal shear of the two spots associated to the above periods has been calculated. We found that the shear of the two spots $\alpha_{in}$ was on average \Malpha\% lower than $\alpha_{out}$. Furthermore, comparing $\alpha_{out}$ to the total equator-to-pole shear we find that $\alpha$ has been underestimated by \totshear\%. Especially, we found that the detection of DR is challenging for stars with an equator-to-pole shear of less than 6\% which usually yields large errors. \\
In our model each light curve is composed by a fixed number of spots rotating at defined latitudes. On the sun the situation is different. Spots vanish or are created while rotation takes place. Their preferred latitudes of occurrence are around $\theta=30\degree$ and during the 11-year activity cycle they migrate towards the equator. So far, our program does not account for spot lifetimes or meridional drifts. \\
The least known model parameter is the number of spots on the surface and their associated size. Alternative to our approach, one could also use more spots with smaller radii keeping the spot filling factor constant, i.e. the fraction of the surface covered with spots. But this also increases the number of available spot periods. Comparing them to the outcome of our analysis bears the risk of choosing a spot that is not responsible for the detected signal. \\
On average the stars in our simulation exhibit 5 spots with radii of $12\degree$ which is due to the chosen parameter range and its uniform distribution. It would be more realistic to couple the spot radii to the activity level of the stars. It is well known that younger stars are more active than e.g. the sun exhibiting on average larger active regions whereas on the sun small spots at preferred latitudes are observed. Our Monte-Carlo simulation does not assign activity levels to the stars which means that DR studies of rather inactive stars like the sun are not fully covered by our model so far. \\
Our method is limited to a relative DR of $\alpha<0.5$. Allowing for $\alpha \ge 0.5$ in the model would lead to confusion between real spot periods and aliases of faster rotating spots with $P1 \leq 1/2 \; P2$. This is a general problem of DR determination from photometric data.
Our lower limit in relation (\ref{limits}) accounts for the frequency resolution in the periodograms. This limit is only relevant to keep reasonable computation time and could in principle be discarded in contrast to the upper limit. Observations of DR cover a wide range of $\alpha$ values. The Doppler Imaging technique is particularly sensitive to small DR limited by the spot lifetimes, $\alpha \lesssim 0.01$, although there are measurements \citep{Donati2003} which yield $\alpha\approx 0.05$ for LQ Hya, and new measurements \citep{Marsden2011} who find values between $0.005\lesssim\alpha\lesssim 0.14$. The Fourier transform method (e.g. \citet{Reiners2003}) is sensitive to $\alpha > 0.1$ and has been used to determine surface shears as large as 50\% for some A-F stars. \\
We conclude that our results confirm the possibility to reliably detect DR from photometric data using a fast tool with relatively simple mathematical assumptions. In real data it will be difficult to correctly interpret the results because of additional effects like spot evolution, pulsations, instrumental effects, and combinations of them. This does not diminish the power of our analytic tool, but its applicability to real data needs to be tested in a large sample of high-quality light curves. We will apply our method to the exquisite sample of the Kepler satellite in a forthcoming publication.

\bibliography{biblothek_v5}
\bibliographystyle{aa}
\acknowledgements TR acknowledges support from the DFG Graduiertenkolleg 1351 \textit{Extrasolar Planets and their Host Stars}. AR acknowledges financial support from the DFG as a Heisenberg Professor under RE 1664/9-1. 

\end{document}